\documentclass[
 reprint,
 amsmath,amssymb,
 aps,
 prl,
 lengthcheck,
]{revtex4}

\usepackage{graphicx}
\usepackage{dcolumn}
\usepackage{bm}
\usepackage{hyperref}

\begin{document}

\title{Magic wavelengths for lattice trapped Rubidium four-level active optical clock}

\author{Zang Xiao-Run}
\affiliation{
Institute of Quantum Electronics, and State Key Laboratory of Advanced Optical Communication System $\&$ Network, \\
School of Electronics Engineering and Computer Science, Peking University, Beijing 100871, People's Republic of China
}
\author{Zhang Tong-Gang}
\affiliation{
Institute of Quantum Electronics, and State Key Laboratory of Advanced Optical Communication System $\&$ Network, \\
School of Electronics Engineering and Computer Science, Peking University, Beijing 100871, People's Republic of China
}
\author{Chen Jing-Biao}
 \email{jbchen@pku.edu.cn}
\affiliation{
Institute of Quantum Electronics, and State Key Laboratory of Advanced Optical Communication System $\&$ Network, \\
School of Electronics Engineering and Computer Science, Peking University, Beijing 100871, People's Republic of China
}

\date{\today}

\begin{abstract}
After pumped from $5s_{1/2}$ ground state to $6p_{1/2}$ state, the population inversion between $6s_{1/2}$ and $5p_{1/2,3/2}$ will be established for Rubidium four-level active optical clock. In this paper, we calculate AC Stark shift due to lattice trapping laser which dominates the frequency shift of clock transition in lattice trapped Rubidium four-level active optical clock. Several magic wavelengths are found that can form desired optical lattice trapping potential. By choosing a proper intensity and linewidth of trapping laser, the fractional frequency uncertainty of clock transition due to AC Stark shift of trapping laser, is estimated to be below 10$^{-18}$.
\begin{description}
\item[PACS numbers]
06.30.Ft, 32.10.Dk, 32.60.+i.
\end{description}
\end{abstract}

\maketitle

Time or frequency precise measurement is extremely important for fundamental science and technology. Optical clock is, undoubtedly, the best time or frequency measurement instrument implemented to date\cite{PhysRevLett.104.070802,Ludlow28032008,Katori.Nature.5.203}. The most accurate optical clock is Aluminium single-ion clock which has achieved a fractional frequency uncertainty of $8.6\times10^{-18}$\cite{PhysRevLett.104.070802}. Although single-ion optical clock possesses high accuracy, its stability is limited according to Allan variance, $\sigma_y(\tau) \propto 1/\sqrt{N}$\cite{Allan.Proc.IEEE.54.221}. Using neutral atoms trapped in optical lattice is one way to break this limitation for it can greatly increase the total number of atoms\cite{PhysRevLett.91.173005}. Recently, a Strontium optical lattice clock with $10^{-16}$ fractional uncertainty has been reported\cite{Ludlow28032008,Katori.Nature.5.203}, using so called "magic wavelength" trapping light to cancel the light shift of the two clock transition levels involved\citep{PhysRevLett.91.173005}. In such optical lattice system, neutral atoms are trapped in optical lattice and the trapped neutral atoms experience frequency shifts which arise from various of mechanics, i.e., trapping light shift, black-body shift, collision shift, first and second order Doppler shift, etc. In the Lamb-Dick regime\cite{LambDick.PhysRev.89.472,PhysRevLett.91.053001}, collision shift and the first and second order Doppler shift are greatly suppressed since their movements are confined. Therefore, most of the time, the first two terms will dominate the frequency shift.

However, both single-ion trapped optical clock and neutral atoms optical lattice clock mentioned above are working in passive mode which is limited by the linewidth of interrogating laser. On the other hand, an active optical clock\cite{1574003,J.B.Chen.ActiveOpticalClock,springerlink:10.1007/s11434-009-0064-z,PhysRevA.78.013846} will output laser directly whose frequency is determined by optical clock transition. Therefore, it can provide two distinguishing qualities, i.e., a super-narrow quantum-limited linewidth and extremely small cavity pulling shift\cite{J.B.Chen.ActiveOpticalClock}. Theoretically, a Strontium clock using thermal atomic beam implies a quantum-limited linewidth of 0.51 Hz and outshines state-of-art narrowest 6.7 Hz of Hg single-ion clock and 1.5 Hz of Strontium optical lattice clock\citep{J.B.Chen.ActiveOpticalClock}. It promises to improve the stability of the best clocks by 2 orders of magnitude, and it can be extended to quenching laser and superradiant laser\cite{PhysRevLett.102.163601,PhysRevA.81.023818,PhysRevA.81.053809,JILA.nature10920}.

\begin{figure}[!htpb]
\includegraphics[scale=.42]{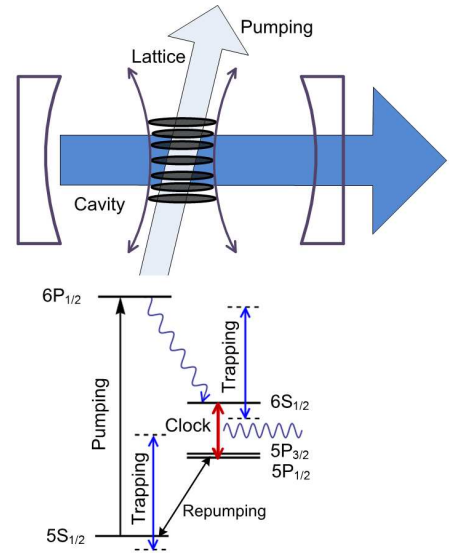}
\caption{\label{fig:RbOpticalLattice2}
Scheme of lattice trapped Rubidium four-level active optical clock, with 421.7 nm pumping light pumps neutral Rubidium from ground state $5s_{1/2}$ to excited state $6p_{1/2}$ and 780 nm, 795 nm repumping light repump atoms back to the optical loop, a population inversion between $6s$ and $5p$ states formed with time evolution. And atoms are trapped in optical lattice clock at magic wavelength where light frequency shift of the clock transition arises from the trapping light is cancelled.
}
\end{figure}

Lattice trapped Rubidium four-level active optical clock (Fig. \ref{fig:RbOpticalLattice2}) is a new experimental scheme to realize active optical clock.
A population inversion exists between the clock transition $6s_{1/2} - 5p_{1/2, 3/2}$ with 8.4\% in $6s_{1/2}$ and 3.3\% in $5p_{3/2}$ when pumping from $5s$ to $6p_{1/2}$ with a laser of 421.7 nm wavelength and 318 mW/cm$^2$ intensity\cite{ztg}. The population inversion is formed after 10$^{-6}$ s, thus a steady photon number will be sustained with a bad cavity to form an active optical clock. 
Considering the Rayleigh scattering and Raman scattering in optical lattice, still a trap lifetime lager than 100 s is achievable\cite{PhysRevLett.82.4204,PhysRevA.63.043403,0256-307X-26-9-090601}, such a long period is sufficient for lattice trapped Rubidium four-level active optical clock. In this paper we derive several magic wavelengths for such a system.

Neutral Rubidium can be cooled and trapped in optical lattice due to dipole gradient force\cite{V.S.Letokhov.book,PhysRevLett.65.33,H.J.Metcalf.book}. The trapping light, meanwhile, causes light shift of energy levels which is affected by both the wavelengths and the intensities of trapping light.

According to quantum perturbation theory, energy shift of arbitrary state can be expressed as follow,
\begin{equation}
\Delta E = -\frac{1}{2}(E)^2\alpha_J^{(2)}
\end{equation}
where E is the electric field strength, $\alpha_J^{(2)}$ is the polarizability of related states. Here, we consider the electric dipole interaction and neglect all higher-order perturbations.

We consider the linearly polarized light to exclude the axial polarizability associated with light polarization and choose the orientation of the trapping light polarization along the magnetic z-axis, finally the polarizability expression writes as\citep{PhysRevA.76.052509},
\begin{equation}
\alpha_J^{(2)} = \alpha_J^{S,(2)} + \frac{3m_j^2-J(J+1)}{J(2J-1)}\alpha_J^{T,(2)}\label{alphaTotal}
\end{equation}
where $\alpha_J^{S,(2)}$ and $\alpha_J^{T,(2)}$ represent scalar and tensor polarizabilities, and all variables and signs follow Ref. \citep{PhysRevA.76.052509}.

All reduced matrix elements (RMEs) and energy levels needed in our calculation are listed in Table \ref{RbEnergyLevel&RME}.

Part of RMEs are derived from the equation 
$
d^2 = 3c^3W(1+2J)/(4\omega ^3)\label{ReducedMatrixElement}
$. 
Where d is reduced matrix element, W is the transition probability, J is the total angular momentum quantum number. Above equation has used atomic units, where $ m_{e}=\hbar=e=\frac{1}{4\pi\epsilon_{0}}=1 $.

To verify the validity of our calculation for rubidium atom we reproduce some calculations from Ref.~\cite{PhysRevA.76.052509}, then we calculate the AC polarizabilities of $5p_{1/2}, 5p_{3/2}$ and $6s_{1/2}$ using Eqs.~(\ref{alphaTotal}).

\begin{table}[!htbp]
\caption{\label{RbEnergyLevel&RME}
Neutral Rubidium energy levels in wavelength (nm) and electric-dipole matrix elements in (a.u.). For $5p_{1/2}, 5p_{3/2}$ states, transitions to $5s, 6s, 7s, 8s$ and to $4d, 5d, 6d$ are included, and for $6s_{1/2}$ state, transitions to $5p, 6p, 7p, 8p$ are included.}
\begin{ruledtabular}
\begin{tabular}{clcrc}
\quad&transition&
wavelength (nm)&
RMEs (a.u.)
&\ \ \ \quad\\
\hline
&$5p_{1/2}-5s_{1/2}$&794.98  \footnotemark[1]& 4.231\footnotemark[1]&\\
&                   &794.979 \footnotemark[2]& 4.221\footnotemark[3]&\\ 
&$5p_{1/2}-6s_{1/2}$&1323.88 \footnotemark[1]& 4.146\footnotemark[1]&\\
&$5p_{1/2}-7s_{1/2}$&728.20  \footnotemark[1]& 0.953\footnotemark[1]&\\
&$5p_{1/2}-8s_{1/2}$&607.24  \footnotemark[1]& 0.502\footnotemark[1]&\\
&$5p_{1/2}-4d_{3/2}$&1475.65 \footnotemark[1]& 8.051\footnotemark[1]&\\
&                   &                        &                      &\\
&$5p_{1/2}-5d_{3/2}$&762.10  \footnotemark[1]& 1.35 \footnotemark[1]&\\
&                   &                        &                      &\\
&$5p_{1/2}-6d_{3/2}$&620.80  \footnotemark[1]& 1.07 \footnotemark[1]&\\
&                   &                        &                      &\\
\hline
&$5p_{3/2}-5s_{1/2}$& 780.24 \footnotemark[1]& 5.977\footnotemark[1]&\\
&                   & 780.241\footnotemark[2]& 5.956\footnotemark[3]&\\
&$5p_{3/2}-6s_{1/2}$& 1366.87\footnotemark[1]& 6.05 \footnotemark[1]&\\
&$5p_{3/2}-7s_{1/2}$& 741    \footnotemark[1]& 1.35 \footnotemark[1]&\\
&$5p_{3/2}-8s_{1/2}$& 616.13 \footnotemark[1]& 0.708\footnotemark[1]&\\
&$5p_{3/2}-4d_{3/2}$& 1529.26\footnotemark[1]& 3.63 \footnotemark[1]&\\
&$5p_{3/2}-4d_{5/2}$& 1529.37\footnotemark[1]& 10.9 \footnotemark[1]&\\
&$5p_{3/2}-5d_{3/2}$& 776.16 \footnotemark[1]& 0.67 \footnotemark[1]&\\
&$5p_{3/2}-5d_{5/2}$& 775.98 \footnotemark[1]& 1.98 \footnotemark[1]&\\
&$5p_{3/2}-6d_{3/2}$& 630.1  \footnotemark[1]& 0.51 \footnotemark[1]&\\
&$5p_{3/2}-6d_{5/2}$& 630.01 \footnotemark[1]& 1.51 \footnotemark[1]&\\
\hline
&$6s_{1/2}-5p_{1/2}$ & 1323.88\footnotemark[2]& 4.119 \footnotemark[3]&\\
&                    &                        &                       &\\
&$6s_{1/2}-6p_{1/2}$ & 2791.29\footnotemark[2]& 9.684 \footnotemark[3]&\\
&$6s_{1/2}-7p_{1/2}$ & 1298.28\footnotemark[2]& 0.999 \footnotemark[3]&\\
&$6s_{1/2}-8p_{1/2}$ & 1030.67\footnotemark[2]& 0.393 \footnotemark[3]&\\
&$6s_{1/2}-5p_{3/2}$ & 1366.87\footnotemark[2]& 6.013 \footnotemark[3]&\\
&                    &                        &                       &\\
&$6s_{1/2}-6p_{3/2}$ & 2732.18\footnotemark[2]& 13.592\footnotemark[3]&\\
&$6s_{1/2}-7p_{3/2}$ & 1292.39\footnotemark[2]& 1.54  \footnotemark[3]&\\
&$6s_{1/2}-8p_{3/2}$ & 1028.67\footnotemark[2]& 0.628 \footnotemark[3]&\\
\end{tabular}
\end{ruledtabular}
\footnotetext[1]{Ref.[\onlinecite{PhysRevA.76.052509}]}
\footnotetext[2]{Ref.[\onlinecite{citeulike:8063389,NIST.Nov.2011.Online}]}
\footnotetext[3]{Ref.[\onlinecite{PhysRevA.69.022509}]}
\end{table}

\begin{figure}[!htbp]
\includegraphics[scale=.06]{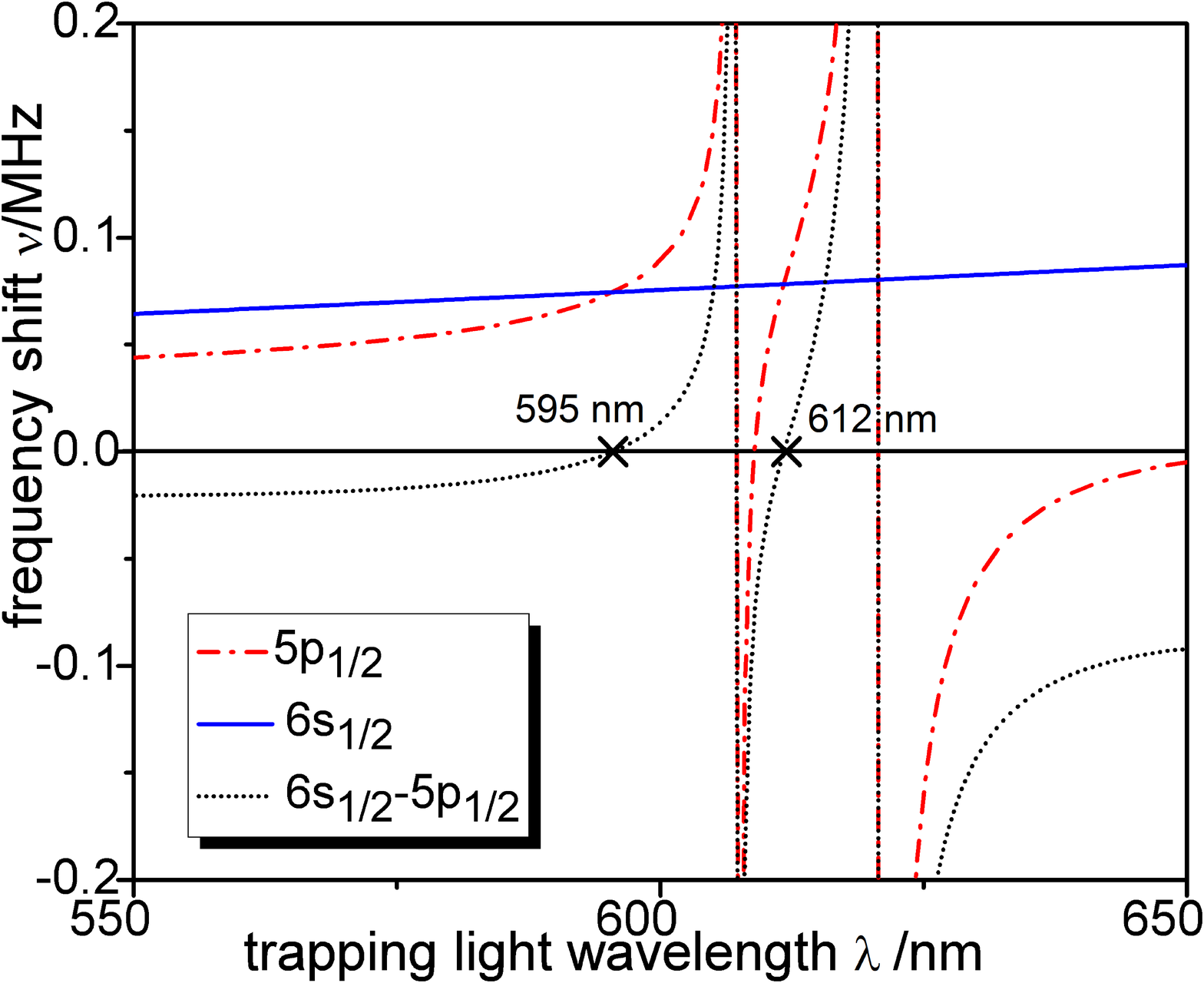}
\caption{\label{fig:FrequencyShift1}
Figure shows, respectively, the frequency shift of the states $5p_{1/2}$ and $6s_{1/2}$. $5p_{1/2}$ is plotted in dash dot line (red), $6s_{1/2}$ is plotted in solid line (blue), and the different frequency shift between them is plotted in short dot line (black). The different frequency shift goes to zero at magic wavelength, which is marked as cross symbol. This figure is plotted with the assumption that the intensity of trapping laser is 10 kW/cm$^2$.}
\end{figure}

The polarizability of $6s_{1/2}$ results from the summation of all the levels that have electric-dipole interactions with $6s_{1/2}$. The transition frequencies $np$-$6s$ is readily to obtain from NIST website\cite{NIST.Nov.2011.Online} and other related papers\cite{citeulike:8063389}, the reduced electric-dipole moment values are from M. S. Safronova et al.\citep{PhysRevA.69.022509} who calculated the reduced electric-dipole moments up to $8p$-$6s$ using a relativistic all-order method. Using these values to calculate the polarizability of $6s$ state is enough for a rough evaluation, and we will discuss the accuracy in the last section.

\begin{figure}[!htbp]
\includegraphics[scale=.06]{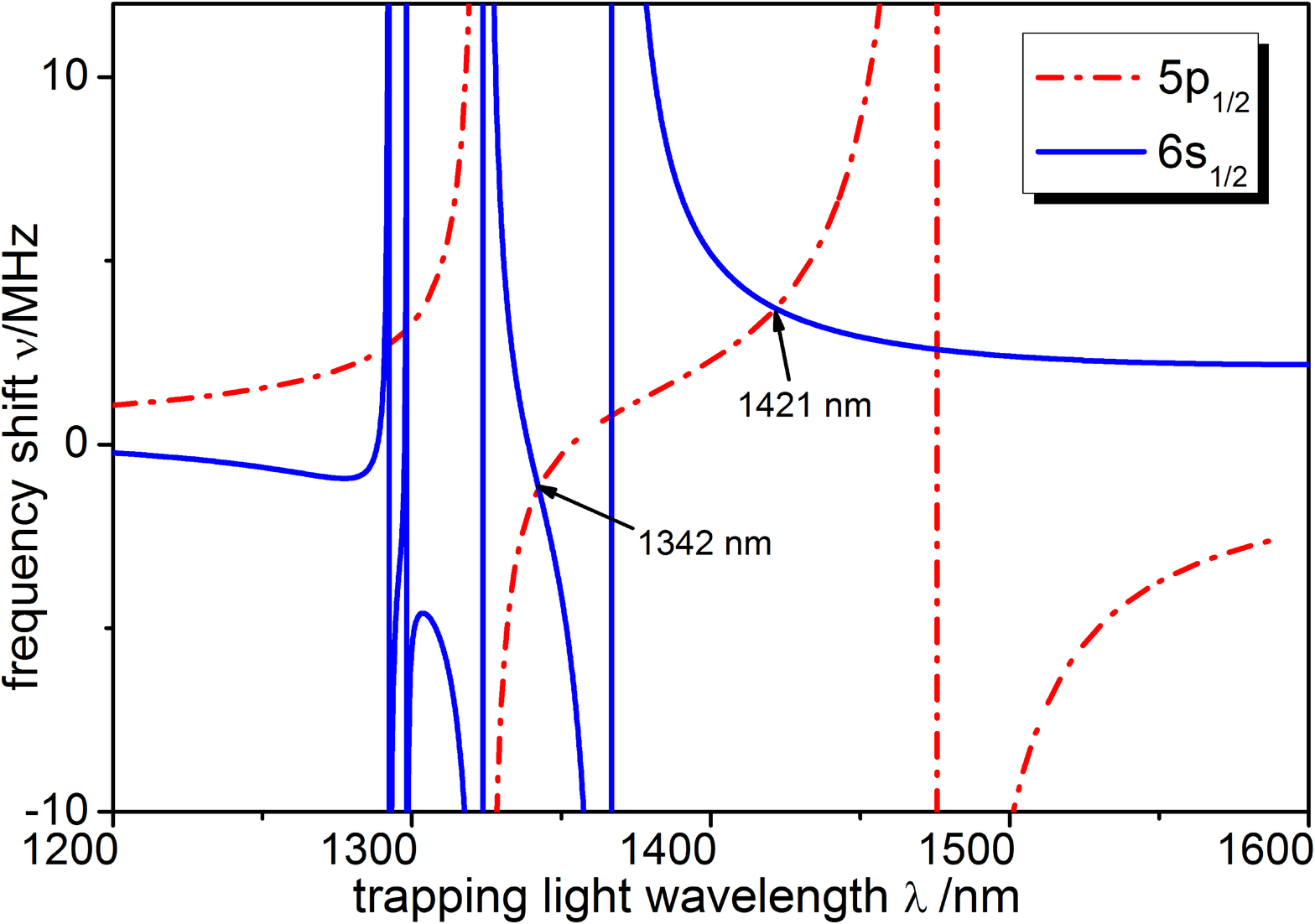}
\caption{\label{fig:FrequencyShift11200-1600nm}
Figure shows, respectively, the frequency shift of the states $5p_{1/2}$ and $6s_{1/2}$. $5p_{1/2}$ is plotted in dash dot line (red), $6s_{1/2}$ is plotted in solid line (blue). This figure is plotted with the assumption that the intensity of trapping laser is 10 kW/cm$^2$.}
\end{figure}

Fig.\ref{fig:FrequencyShift1} and Fig.\ref{fig:FrequencyShift11200-1600nm} show the difference frequency shift between $6s_{1/2}$ and $5p_{1/2}$ levels while the trapping light frequency varies from 550 nm to 650 nm, 1200 nm to 1600 nm. The light shift of $5p_{1/2}$ is dominated by the electric dipole interaction $5p_{1/2}$-$8s_{1/2}$ at wavelength 595 nm which is lower than the resonance wavelength. At wavelength 612 nm, $5p_{1/2}$ is greatly shifted by two electric dipole interaction $5p_{1/2}$-$8s_{1/2}$ and $5p_{1/2}$-$6d_{3/2}$. Two magic wavelengths, 1342 nm and 1421 nm, are found between 1200 nm and 1600 nm.

\begin{figure}[!htbp]
\includegraphics[scale=.06]{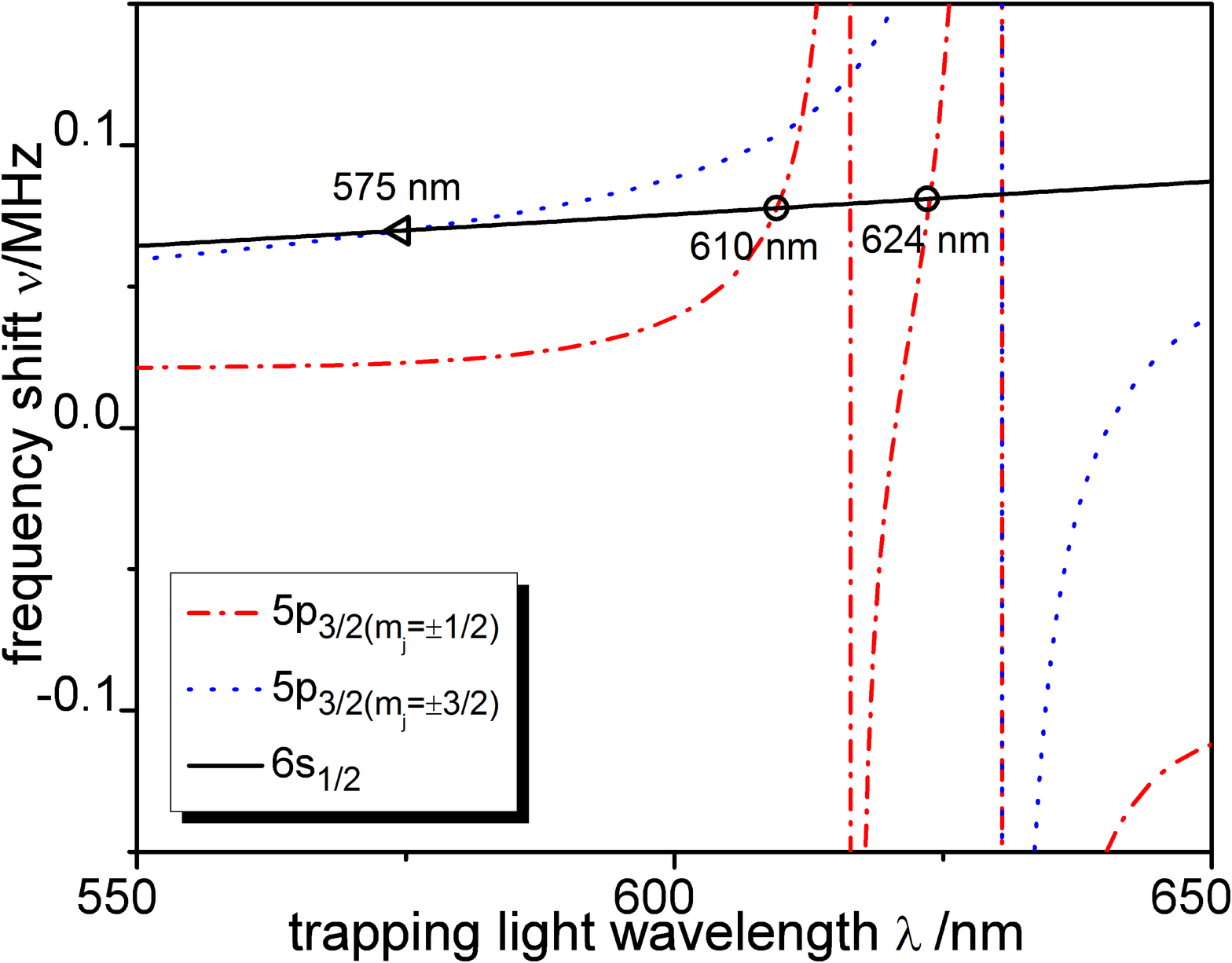}
\caption{\label{fig:FrequencyShift2}
Figure shows, respectively, the frequency shift of the states $5p_{3/2}(m_j=\pm\frac{1}{2})$, $5p_{3/2}(m_j=\pm\frac{3}{2})$ and $6s_{1/2}$. $5p_{3/2}(m_j=\pm\frac{1}{2})$ is plotted in dash dot line (red), $5p_{3/2}(m_j=\pm\frac{3}{2})$ is plotted in dot line (blue), and $6s_{1/2}$ is plotted in solid line (black). Light shift of the states $5p_{3/2}(m_j=\pm\frac{1}{2})$, and $6s_{1/2}$ intersect at the magic wavelength which is marked as circle symbol. Triangle symbol marks the cross point of light frequency shift of states $5p_{3/2}(m_j=\pm\frac{3}{2})$ and $6s_{1/2}$. This figure is plotted with the assumption that the intensity of trapping laser is 10 kW/cm$^2$.}
\end{figure}

\begin{figure}[!htbp]
\includegraphics[scale=.06]{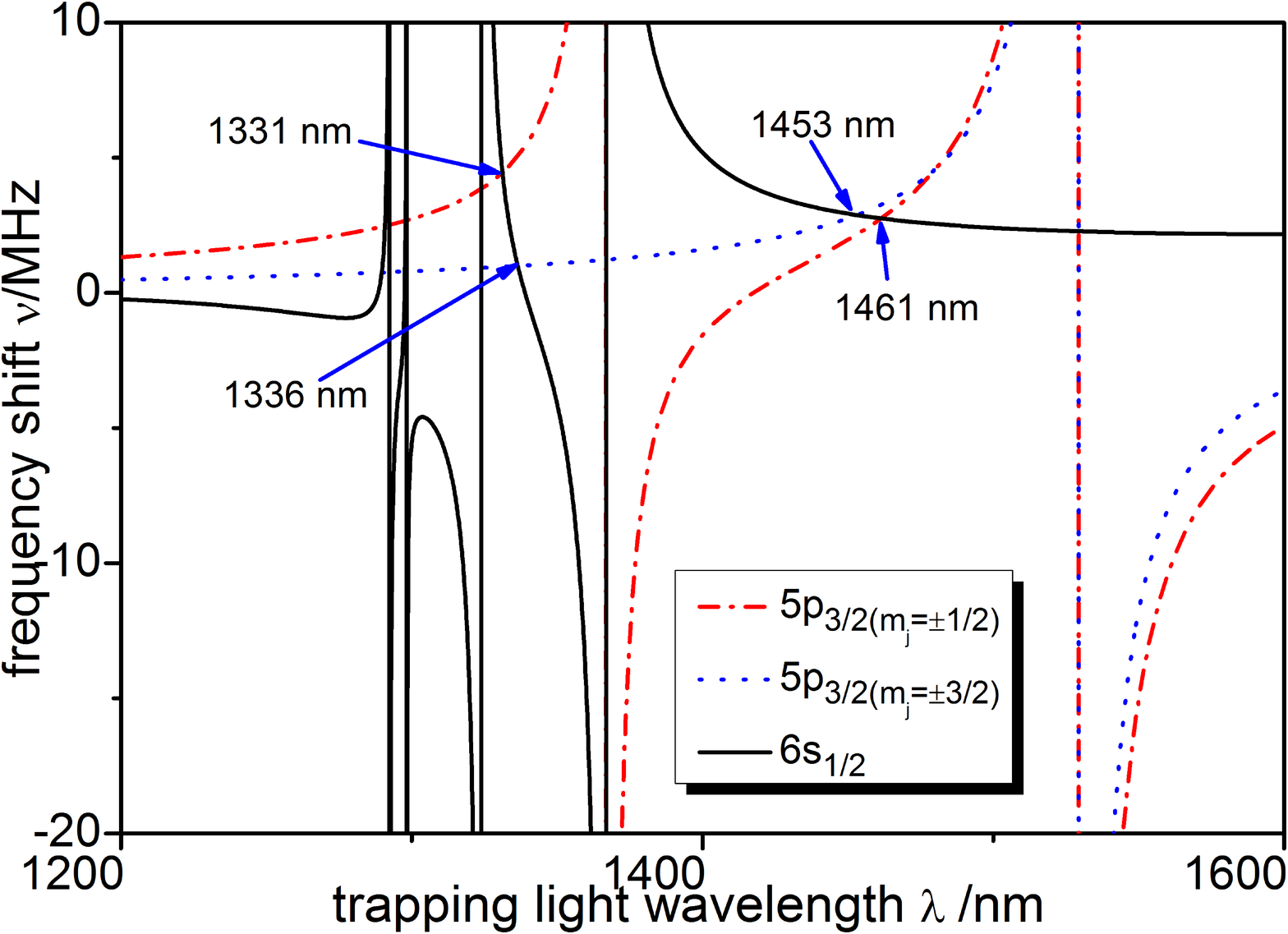}
\caption{\label{fig:FrequencyShift21200-1600nm}
Figure shows, respectively, the frequency shift of the states $5p_{3/2}(m_j=\pm\frac{1}{2})$, $5p_{3/2}(m_j=\pm\frac{3}{2})$ and $6s_{1/2}$. $5p_{3/2}(m_j=\pm\frac{1}{2})$ is plotted in dash dot line (red), $5p_{3/2}(m_j=\pm\frac{3}{2})$ is plotted in dot line (blue), and $6s_{1/2}$ is plotted in solid line (black). Light shift of the states $5p_{3/2}(m_j=\pm\frac{1}{2})$, and $6s_{1/2}$ intersect at the magic wavelength which is marked as circle symbol. Triangle symbol marks the cross point of light frequency shift of states $5p_{3/2}(m_j=\pm\frac{3}{2})$ and $6s_{1/2}$. This figure is plotted with the assumption that the intensity of trapping laser is 10 kW/cm$^2$.}
\end{figure}

Fig.\ref{fig:FrequencyShift2} and Fig.\ref{fig:FrequencyShift21200-1600nm} show the difference frequency shift between $6s_{1/2}$ and $5p_{3/2}$ levels while the trapping light frequency varies from 550 nm to 650 nm, 1200 nm to 1600 nm. With respect to state $5p_{3/2}(|m_j|=\frac{1}{2})$, the wavelength 610 nm is lower than the resonance $5p_{3/2}$-$8s_{1/2}$ wavelength, while wavelength 624 nm is between the resonance $5p_{3/2}$-$8s_{1/2}$ and resonance $5p_{3/2}$-$6d_{3/2}$. For state $5p_{3/2}(|m_j|=\frac{3}{2})$, the wavelength 575 nm lies on the blue side of the resonance $5p_{3/2}$-$6d_{3/2, 5/2}$. For optical clock transition $6s_{1/2}$-$5p_{3/2}(|m_j|=\frac{1}{2})$, 1331 nm and 1461 nm are magic wavelengths; for $6s_{1/2}$-$5p_{3/2}(|m_j|=\frac{1}{2})$, two magic wavelengths 1336 nm and 1453 nm are found.

Magic wavelengths in the range of 550 nm to 650 nm and 1200 nm to 1600 nm are listed in Table\ref{MagicWavelengths}. The frequency uncertainties at each magic wavelengths are listed in Table\ref{Uncertainty} by properly adjusting trapping laser intensities, i.e., trapping depths.

\begin{table}[!htbp]
\caption{\label{MagicWavelengths}
The magic wavelengths for optical clock transitions $6s_{1/2}$-$5p_{1/2}$ and $6s_{1/2}$-$5p_{3/2}$ and lattice trapping depths to the intensity of lattice laser $\Delta\nu/I$ (kHz/kW/cm$^2$) refer to those magic wavelengths are listed in this table.}
\begin{tabular}{cc|cc}
\hline\hline
\multicolumn{2}{c|}{$6s_{1/2} - 5p_{1/2}$}&
\multicolumn{2}{c}{$6s_{1/2} - 5p_{3/2}$}\\
$\lambda$ (nm)&
$\Delta\nu/I$ (kHz/kW/cm$^2$)&
$\lambda$ (nm)&
$\Delta\nu/I$ (kHz/kW/cm$^2$)\\
\hline
595		&$-74.5$	&575	&$-69.7$\\
612		&$-78.2$	&610	&$-77.7$\\
1342	&$1196.7$	&624	&$-81.0$\\
1421	&$-3707.2$	&1331	&$-4453.4$\\
		&			&1336	&$-995.2$\\
		&			&1453	&$-2878.1$\\
		&			&1461	&$-2753.4$\\		
\hline\hline
\end{tabular}
\end{table}

The reason why choosing four-level quantum system rather than three-level, optical lattice clock rather than thermal atomic beam are as follow. As for thermal atomic beam clock, residual Doppler shift will be extremely large, however optical lattice clock remove that effect otherwise since atoms are  laser cooled and trapped in Lamb-Dick regime. On the other hand, three-level system is affected by light shift of pumping laser. Four-level quantum system, however, can avoid such disadvantage by choosing clock transition without involving ground state which is connected to pumping laser directly.

In our calculation, we used the reduced matrix element data up to $8p$-$6s$ transition. Since the frequency shift of $6s$ due to $8p_{1/2}$ and $8p_{3/2}$ states contribute less than $0.081\%$, at all magic wavelengths, to the total polarizability of $6s$, therefore we believe that states above $8p$ will contribute fewer. As for the polarizability of $5p_{1/2}$, states $5s_{1/2}$, $6s_{1/2}$ and $4d_{3/2}$ will dominate, states $8s_{1/2}$ and $6d_{3/2}$ totally have less than $0.639\%$ effect.

In conclusion, four magic wavelengths 595 nm, 612 nm, 1342 nm and 1421 nm are available for transition $6s_{1/2}$-$5p_{1/2}$, seven magic wavelengths 575 nm, 610 nm, 624 nm, 1331 nm, 1336 nm, 1453 nm and 1461 nm are candidates for transition $6s_{1/2}$-$5p_{3/2}$ (listed in Table\ref{MagicWavelengths}). To obtain a 10$^{-18}$ fractional uncertainty or even less, one must well control the linewidth of lattice laser, i.e., below 10 kHz as listed in Table\ref{Uncertainty}. With respect to magic wavelengths between 550 nm and 650 nm, a further discussion is needed since those wavelengths are going to above the ionization limit of $6s_{1/2}$ (741 nm). In addition, dynamic stark shift induced by black-body radiation will be discussed elsewhere.

\begin{table}[!htbp]
\caption{\label{Uncertainty}
Adjusting the intensities $I$ (W/cm$^2$) of trapping laser to provide sufficient trapping depths $\Delta\nu$ (kHz), for each magic wavelengths $\lambda$ (nm) the frequency uncertainties (mHz) due to lattice laser frequency fluctuation (assumed to be 10 kHz) are listed in this table.}
\begin{tabular}{lccc}
\hline\hline
$\lambda$ (nm)&
$I$ (W/cm$^2$)&
$\Delta\nu$ (kHz)&
uncertainty (mHz)\\
\hline
595		&1E4	&$-745$		&4.3E-4\\
612		&1E4	&$-782$		&4.6E-4\\
1342	&1E3	&$1196.7$	&3.2E-1\\
1421	&200	&$-741.4$	&9.0E-3\\
\hline
575		&1E4	&$-697$		&3.9E-4\\
610		&1E4	&$-777$		&4.6E-4\\
624		&1E4	&$-810$		&4.8E-4\\
1331	&200	&$-890.7$	&2.0E-1\\
1336	&1E3	&$-995.2$	&4.2E-1\\
1453	&250	&$-863.4$	&4.6E-3\\
1461	&250	&$-688.4$	&3.8E-3\\
\hline\hline
\end{tabular}
\end{table}

Moreover, the pumping light of wavelength 421.7 nm and intensity 318 mW/cm$^2$ contributing frequency corrections to optical clock transitions of $6s_{1/2}$-$5p_{1/2}$, $6s_{1/2}$-$5p_{3/2(m_j=\pm\frac{1}{2})}$, $6s_{1/2}$-$5p_{3/2(m_j=\pm\frac{3}{2})}$ are 0.45 Hz, 0.65 Hz and 0.28 Hz each and related uncertainties will be no more than 1 mHz when an order of 10$^{-3}$ intensity fluctuation of pumping laser is considered. Although our results for Rubidium atom is specified, it is ready to be generalized to other alkaline atoms.

This work was initiated by the National Natural Science Foundation of China (NSFC)(grants 10874009 and 11074011). The authors thank Xiaoji Zhou, Xia Xu and Anpei Ye for discussions.\\\\
\begin{Large}
\textbf{References}
\end{Large}

\end{document}